\documentclass[preprint,12pt]{elsarticle}

\usepackage{amssymb}
 \usepackage{amsthm}
\usepackage{amsmath}
\usepackage{dcolumn}
\usepackage{bm}
\usepackage{stfloats}
\usepackage[caption=false,font=footnotesize]{subfig}
\usepackage{dutchcal,mathrsfs}
\journal{Physica A}

\begin{document}

\begin{frontmatter}



\title{Spectral Clustering Methods for Multiplex Networks}


\author[label1]{Daryl R. DeFord}
\ead{ddeford@math.dartmouth.edu}
\ead[url]{math.dartmouth.edu/~ddeford}
\author[label2]{Scott D. Pauls}
\ead{scott.d.pauls@dartmouth.edu}
\ead[url]{math.dartmouth.edu/~pauls}

\address[label1]{Department of Mathematics, Dartmouth College, Hanover, NH, USA}

\address[label2]{Department of Mathematics, Dartmouth College, Hanover, NH, USA}

\begin{abstract}
Multiplex networks offer an important tool for the study of complex systems and extending techniques originally designed for single--layer networks is an important area of study. One of the most important methods for analyzing networks is clustering the nodes into communities that represent common connectivity patterns. In this paper we extend  spectral clustering to multiplex structures and discuss some of the difficulties that arise in attempting to define a natural generalization. In order to analyze our approach, we describe three simple, synthetic multiplex networks and compare the performance of different multiplex models. Our results suggest that a dynamically motivated model is more successful than a structurally motivated model in discovering the appropriate communities. 
\end{abstract}


\end{frontmatter}



\section{Introduction}
Clustering the nodes of networks is an important tool for the dimension reduction of models of complex systems.  For single layer networks, numerous methods and approaches appear in the literature –-- $k$-means or $k$-nearest neighbor, modularity maximization, and spectral clustering, among others.  Our goal in this work is to provide a generalization of spectral clustering to multiplex networks while illuminating the difficulties and choices that must be made when transporting these ideas to a multiplex network.
In the setting of a single layer network, $(N,E)$, where $N$ is the node set and $E$ is the edge set, spectral clustering is directly related to a geometric problem on the network –-- how do we cut the network into pieces while doing the least amount of damage?  In the base case, where we cut the network into two pieces with node sets $N_1$ and $N_2$, the damage is simply the sum of the weights of the edges between nodes in $N_1$ and those in $N_2$:
\[cut(N_1,N_2 )=  \sum_{ i \in N_1, j \in N_2} w(i,j),\]
where $w(i,j)$ give the weight of the connection between $i$ and $j$. Minimizing the damage over all possible partitions into two subsets, while an $NP$-hard problem, relaxes to the more tractable problem of finding an eigenvector associated to the first non-trivial eigenvalue of the graph Laplacian associated to the network \cite{shi_normalized_2000, white_spectral_2005,  newman1, newman2, vonLuxberg}, providing a link between the cut problem and the evolution of the diffusion dynamics on the network.

Transporting this basic idea of partitioning to a multiplex network is not trivial and requires choices --- none of which have obvious forms and all of which can have significant impact on the outcome.  We focus first on a question of fundamental importance concerning the nature of the nodes.  Consider the fixed node set $N$, and a multiplex network with $k$ edges sets $\{E_1,\dots,E_k\}$ corresponding to the layers of the network.  When dividing up our node set, we must first decide whether the nodes themselves are indivisible or not.  In many cases, nodes are single entities –-- like people in a social network --– which can be separated from one another but not split into component parts.  In other situations, nodes are aggregations of entities which act in concert –-- for example countries that trade with one another comprise many individuals, cartels, and firms --– but where it may be plausible that different components separate from one another upon clustering. This distinction is crucial; at the extremes we can reduce the problem to an associated single layer network problem while in intermediate cases, truly multiplex approaches are necessary.

At one extreme, if we think about cutting the node set into two subsets, $N_1$ and $N_2$, there are many possible ways to measure the damage done which amounts to assigning measures of importance to edges on different layers.  This is equivalent to a choice of aggregation --– edges are collapsed across the layers resulting in a new single layer network associated to the clustering problem, reducing the problem to the single layer clustering problem.  While the mechanics at this point are known, the process to get there is not:  such an aggregation takes different forms depending on the question which drives the clustering. 

Upon relaxing this requirement and allowing nodes to break apart into components and separate into different subsets under partitioning, we must contend with the subtleties of measuring the damage done by such a cut.  What is the cost of cutting a component of a node away from the other components?  What is the cost of cutting an edge on a particular layer?  Answers to such questions are again neither obvious or canonical.  At the other extreme, nodes are very weak aggregate entities and dividing nodes into pieces carries almost no cost.  The limiting case of this simply treats copies of the nodes independently and reduces to a clustering problem on a larger single layer network, one where the node set is $k$ copies of $N$ and the layer edges populate this larger single layer network.  In between these extremes, we have a regime in which choices are delicate and have significant impact on the outcome.

So how do we make these choices?  In a given specific empirical situation, information about the system in question or background theory may determine the cohesiveness of the nodes and the meaningfulness of the edges.  But, as we wish to approach this question theoretically, we are at a loss.  Moreover, thinking in analogue of the single layer case, we might expect a connection to a process akin to diffusion on a multiplex network.  However, for multiplex networks, several candidates for such a process are each potentially appropriate in different situations.

To help see our way through this conundrum, we put forward test cases on which to evaluate candidates for the resolution of the cut problem.  First, we construct another family of models with Erd\H{o}s-R\'enyi networks for each of the layers.   With random layers, we expect no clustering at all.  Consequently, we expect the node copies, however weakly associated to one another, to group together.  Second, we construct a family of networks which have the same fixed community structure on each layer using stochastic block models (SBM) \cite{SBM}.  In this case, any viable method should reveal these communities.  Third, we create a multiplex network with different community structures on each layer again using the SBM.  Here, we expect the clustering to depend on the relative importance of the layers themselves leading to different partitions based on different choices.

For each of these three families, we perform spectral clustering using two different formulations of a multiplex graph Laplacian associated to the cut problem.  
First, we test the popular supra-adjacency structural model \cite{Gomez} which leads to an analysis of the supra-Laplacian for spectral clustering.  
Broadly, we find that the supra-adjacency representation of the multiplex network does well in some experiments, but fails in others.  In particular, it cannot reliably find the fixed communities in the second case unless the inter-layer connectivity is sufficiently strong.  Second, we use a structural model based on a different multiplex Laplacian dynamical model \cite{JCN} and construct the associated graph Laplacian for spectral clustering.    This multiplex dynamic model  fares better across all the experiments.  In our first two tests, this operator performs as we expect, clustering most node copies together when the layers are random graphs, and finding the fixed communities exactly when the layers encode the same fixed community.  

But it is in the third set of experiments where operators show the most substantial differences.  For a large range of inter-layer weights, spectral clustering using the supra-Laplacian finds four clusters,  splitting the layers from one another and detecting the fixed communities in each layer.  Results of spectral clustering using the second model are more complex.  An exploration of the model parameter space for two layer networks with overlapping SBM clusters shows that this formulation of spectral clustering reveals partitions of the multiplex network that subtly interpolate between the layer partitions depending on the relative strengths of the importance of the layers.  These results demonstrate an area where this version of multiplex spectral clustering is more than a simple extension of single layer spectral clustering.  In that case, the choices of the parameters in the model allow us to encode features of the multiplex network which reflect the underlying investigation. One formulation of this is to consider the following question:  ``Are there layers which have more relevance to certain questions?''  If so, use appropriate choices of model parameters to emphasize those layers and de-emphasize the rest.  

\section{Spectral clustering for single layer networks}
Before diving into clustering for multiplex networks, we briefly review spectral clustering for single layer networks to both set up our terminology and to point out a subtlety of the technique which is useful when considering undirected networks.  If $A$ is the adjacency matrix for a single layer network, we define a cut of the network into two pieces, $\mathscr{B}$ and $\mathscr{C}$ using an {\em indicator vector},
\begin{equation}
\begin{split}
s(i) =
\begin{cases}
1 \;\;\text{if $i \in \mathscr{B}$}\\
-1 \;\;\text{if $ i \in \mathscr{C}$}
\end{cases}
\end{split}
\end{equation}
Then, we calculate the damage done by performing the cut by summing the weights of the edges that must be deleted to achieve the separation of $\mathscr{B}$ from $\mathscr{C}$,
\[cut(\mathscr{B},\mathscr{C}) = \sum_{p \in \mathscr{B}, q \in \mathscr{C}} A_{pq} = \sum_{p,q = 1}^n A_{pq}\frac{1}{2}(1-s(p)s(q)).\]
We can realize this measurement in terms of the graph Laplacian:
\begin{equation}
\begin{split}
cut(\mathscr{B},\mathscr{C}) &=\sum_{i,j = 1}^n A_{ij}\frac{1}{2}(1-s(i)s(j)) \\
&= \frac{1}{2} \left (\sum_{i,j=1}^n A_{ij} - \sum_{i,j=1}^n s(i) A_{ij}s(j) \right ) \\
\intertext{If we sum the first term over $j$, we get the sum of in-degrees of the network while if we sum first over $i$, we get the sum of the out-degrees.  To facilitate using directed networks, we make use of a symmetrization and rewrite this sum as the sum of the average of the two degrees, letting $deg(i)=\frac{1}{2}(deg_{in}(i)+deg_{out}(i))$.  Similarly, as $s^TAs=(s^TAs)^T=s^TA^Ts$, we can replace the second summand with $\frac{1}{2}(s(i) (A_{ij}+A_{ji})s(j))$.}
cut(\mathscr{B},\mathscr{C}) &= \frac{1}{2} \left (\sum_{i=1}^n deg(i) - \frac{1}{2}\sum_{i,j=1}^n s(i) (A_{ij}+A_{ji})s(j) \right ) \\
&= \frac{1}{2} \bigg (\sum_{i,j=1}^n s(i)deg(i)s(j)\delta_{ij}- \frac{1}{2}\sum_{i,j=1}^n s(i)(A_{ij}+A_{ji})s(j) \bigg ) \\
&= \frac{1}{2} (s^T Ds - \frac{1}{2}s^T (A+A^T) s) = \frac{1}{2} s^TLs
\end{split}
\end{equation}
To emphasize the choice above, $D$ is the diagonal matrix with $D_{ii}=\frac{1}{2}(deg_{in}(i)+deg_{out}(i))$ and hence $L=D-\frac{1}{2}(A+A^T)$ is the graph Laplacian associated with the network given by the symmetrization of $A$.  Minimizing the cut is then equivalent to minimizing $s^TLs$ over all $2^n$ possible indicator vectors.  This problem is NP-hard, but we can relax the problem by allowing $s$ to take real values.  The resulting optimization can be solved exactly by taking $s$ to be an eigenvector associated to the smallest non-zero eigenvalue of $L$.  The standard approach generating the two clusters from this vector is to separate the nodes by the sign of the entries of the eigenvector.

\section{Previous work}
The clustering problem in single layer networks has a long history (e.g. see the survey article \cite{schaeffer_survey:_2007}) and many initial approaches to multiplex clustering begin by aggregating the network to a single layer and applying one of the standard techniques. However, this approach forces all copies of each node to belong to the same cluster, which is not always desirable, as we discussed above. At the same time many multiplex networks have significant correlations in layer connectivity and thus some amount of aggregation is desirable to avoid over--modeling \cite{domenico_structural_2015, taylor_enhanced_2016, taylor_detectability_2016}.

Although the particular problem of generalizing spectral clustering to multiplex networks does not seem to have generated much attention in the literature, other single layer techniques have been generalized to the multiplex setting. The most commonly used approaches are based on measures of modularity \cite{Girvan11062002, newman_modularity_2006} which defines communities as subsets where the intra--cluster connections are more dense than would be expected at random. These approaches make explicit use of a null model, usually the  Erd\"{o}s--Reny\`{i} or configuration models.

In \cite{mucha_community_2010} Mucha et al. developed a general framework adapting modularity methods to multiplex networks by characterizing an associated stochastic process similar to the Markov stability objective function developed in \cite{lambiotte_random_2014, delvenne_stability_2010, lambiotte_random_2014-1}.  A careful analysis of the local structure of the communities detected by these random walk based measures is presented in \cite{jeub_local_2015}. Another extension of modularity to the multiplex setting is developed in \cite{bennett_detection_2015} from a computational perspective. In \cite{didier_identifying_2015} Didier et al. compare this approach to modularity clustering on aggregates of multiplex data and show that the intrinsically multiplex methods outperform aggregation. Recently, \cite{paul_null_2016} develops several different possible null models for network modularity from a statistical perspective. 

In addition to modularity based methods, several versions of the SBM have been defined for multiplex networks. A detailed statistical analysis of asymptotic consistency of spectral clustering and the MLE for recovering communities in multiplex networks drawn from a SBM is presented in \cite{han_consistent_2014}. In \cite{peixoto_inferring_2015}, several mutlilayer SBM versions were introduced along with a Bayesian estimation process for the parameters. Restricted SBM versions for multilayer networks were introduced in \cite{paul_community_2015} and extended to degree corrected versions in \cite{paul_null_2016}. These types of models were also used by Taylor et al. in their studies of communities and layer aggregation \cite{taylor_enhanced_2016, taylor_detectability_2016}.

In addition to methods developed from network theory there are also general techniques for non--negative tensor factorization applied to multiplex networks. An approach based on higher order random walks is developed in  \cite{wu_general_2016}. A state of the art computational approach is presented in \cite{gligorijevic_non-negative_2016} with with many examples on both synthetic and real--world data. 
Finally, a recent preprint \cite{de_bacco_community_2017} combines the techniques of many of the above methods to develop a tensor factorization method for the multilayer mixed--membership SBM together with a detailed statistical framework for expectation maximization.

\section{The cut problem for multiplex networks}
A multiplex network comprises a node set $N$ and $k$ edge sets, $\{E_1,\dots, E_k\}$, one for each layer.  An element of an edge set, $w^\alpha(i,j) \in E_\alpha$ denotes a directed weighted edge on layer $\alpha$ from node $j$ to node $i$.  Consequently, for layer $\alpha$, we define an adjacency matrix $A^\alpha$ with $A^\alpha_{ij}=w^\alpha(i,j)$.  To facilitate looking at both indivisible and aggregate nodes, we let $\bar{N}$ be the collection of node copies across all layers.  An element $i^\alpha \in \bar{N}$ denotes the copy of node $i$ on layer $\alpha$.

To formulate the cut problem in this setting, we follow the principles of the single layer case --- the damage done by cutting one node from another is the total weight of the edges connecting the two nodes.  So, if $\mathscr{B}$ and $\mathscr{C}$ form a partition of $\bar{N}$, then
\[cut(\mathscr{B},\mathscr{C})= \sum_{i^\alpha \in \mathscr{B}, j^\beta \in \mathscr{C}} W(i^\alpha,j^\beta),\]
where $W$ is the overall weight function.  From our setup, if $\alpha=\beta$, then we have the natural definition $W(i^\alpha,j^\beta)=w^\alpha(i,j)+w^\alpha(j,i)$.  If $\alpha \neq \beta$, we do not have a clear choice of the weight function $W$.  Different approaches to structurally representing multiplex networks lead to different choices of this function.  We focus on two, considering the supra-adjacency representation \cite{Gomez} and a representation based on models of multiplex dynamics \cite{JCN}.  

The supra-adjacency formulation places a clique of intra-layer edges connecting all copies of one node to one another with fixed weight $\mathcal{w}$.  For this case, we have
\begin{equation*}
W(i^\alpha,j^\beta)=
\begin{cases}
w^\alpha(i,j)+w^\alpha(j,i) \;\;\text{if $\alpha=\beta$}\\
\mathcal{w} \;\;\text{if $i=j$ and $\alpha \neq \beta$}\\
0 \;\;\text{otherwise}
\end{cases}
\end{equation*}

The multiplex dynamical model \cite{JCN} uses the same structural representation for the layers as the supra-adjacency formulation --- we represent layer $\alpha$ using the adjacency matrix $A^\alpha$.  However, inter-layer interactions are encoded differently via an $nk \times nk$ block matrix $\mathcal{A}$ where the block are $n \times n$ matrices indexed by the layers.  The $(\alpha,\beta)$ block is given by $C^{\alpha,\beta}A^{\beta}$ where $C^{\alpha,\beta}$ is a diagonal matrix with $C^{\alpha,\beta}_{ii}=m^{\alpha,\beta}_ic^{\alpha,\beta}_i$.  These two constants together model the transfer from $i^{\beta}$ to layer $\alpha$.   

Consequently, with this representation the combined connections between $i^\alpha$ and $j^\beta$ are
\[W(i^\alpha,j^\beta)= m^{\alpha,\beta}_ic^{\alpha,\beta}_iw^\beta(i,j)+m^{\beta,\alpha}_jc^{\beta,\alpha}_j w^\alpha(j,i). \]
In both of these cases,  we can use the calculation of weights to define the analogue of an adjacency matrix, which we denote $\mathfrak{A}$, so that 
\begin{equation}\label{eqn:cut}
cut(\mathscr{B},\mathscr{C})=\bar{s}^T \mathcal{L}_\mathfrak{A} \bar{s},
 \end{equation}
 where $\bar{s}$ is an indicator vector and $\mathcal{L}_\mathfrak{A}$ is the classical graph Laplacian associated to $\mathfrak{A}$.  For the supra-adjacency formulation, we can take the symmetrized supra-adjacency matrix for $\mathfrak{A}$ while for the multiplex dynamic formulation, we can use the symmetrization $\frac{1}{2} \left (\mathcal{A}+\mathcal{A}^T \right )$ for $\mathfrak{A}$.

As with the single layer formulation, symmetrization does not change the value of $cut(\mathscr{B},\mathscr{C})$ but produces an operator which is easily diagonalizable when we solve the relaxed version of the problem.   

\section{Indivisible nodes and disjoint nodes}
Imposing the extra condition that all the copies of a node copies be in the same part of the partition reduces the problem to clustering on a single layer network.  We can summarize the condition by requiring $\bar{s}(i^\alpha)=\bar{s}(i^\beta)$ for all $\alpha,\beta$ for each node $i$. Then, we reformulate the indicator vector as $\mathfrak{I} s$, where $\mathfrak{I}=\begin{pmatrix} I & \dots & I \end{pmatrix}^T$, $I$ is the $n \times n$ identity matrix and $s:N  \rightarrow \{\pm 1\}$ is an indicator vector on $N$.  Then equation \eqref{eqn:cut} reduces to
\[\bar{s}^T \mathcal{L}_\mathfrak{A} \bar{s}= s^T \left ( \mathfrak{I}^T\mathcal{L}_\mathfrak{A} \mathfrak{I} \right ) s. \]
The matrix in parentheses is an $n \times n$ matrix which we view as a reduction of the full multiplex operator.

For the supra-Laplacian formulation above, this matrix is simply the graph Laplacian of the aggregate of the symmetrized layer adjacency matrices $\sum_\alpha \frac{1}{2} (A^\alpha+(A^\alpha)^T)$.  For the dynamic multiplex formulation we have
\[\mathfrak{A} = \left ( \frac{1}{2}(C^{\alpha,\beta} A^\beta +(C^{\beta,\alpha}A^\alpha)^T )\right ),\]
and
\[\mathcal{L}_\mathfrak{A} = D-\mathfrak{A},\]
where $D$ is the diagonal matrix with $D_{ii}$ is the $i^\text{th}$ column sum of $\mathfrak{A}$.  Multiplying on the left and right $\mathfrak{I}^T$ and $\mathfrak{I}$ and its transpose yields the graph Laplacian associated to the single layer network on $N$ with adjacency matrix given by
\[A= \frac{1}{2}\sum_{\alpha,\beta} \left ( C^{\alpha,\beta} A^\beta +(C^{\beta,\alpha}A^\alpha)^T \right ).\]
Note that in the special case where the $C^{\alpha,\beta}=I$, this reduces to the case of the aggregation of the layer adjacency matrices.  

In contrast to the case where we bind all node copies together, if we treat the node copies as distinct entities with no particular relation to one another our clustering problem reduces to clustering on each of the layers.  To encode this in the two formulations above, we select weights to reduce the cost of separating node copies from one another as much as possible.  For the supra-adjacency matrix, we simply pick $\mathcal{w}=0$ while in the multiplex dynamics framework, we pick $m^{\alpha,\alpha}_i =c^{\alpha,\alpha}_i=1$ for all $\alpha,\beta,i$, and $c^{\alpha,\beta}_i=0$ when $\alpha\neq \beta$. With these choices, $\mathfrak{A}$ is the diagonal block matrix with the $\{A^\alpha\}$ on the diagonal in both cases.  Executing spectral clustering with this adjacency matrix then reduces to performing spectral clustering on the layers.  This case is, in a sense, degenerate as there are many zero eigenvalues of the graph Laplacian which correspond to separating groups of layers from one another --- as there are no cross layer connections, there is no cost to creating such cuts.
\section{General Case}
Analyzing the intermediate cases is more subtle due to the interactions between and within the layers. While we take up a more detailed analysis of examples in the subsequent sections, we discuss some initial results here.  Using the supra-Laplacian with a general indicator vector $\bar{s}=(s_1,s_2,\dots,s_k)$, we have 
\begin{equation*}
\begin{split}
\bar{s}^T &\mathcal{L}_\mathfrak{A} \bar{s} \\
&= \bar{s}^T \begin{pmatrix} \mathcal{L}_{\frac{1}{2}(A^1+(A^1)^T)}+(k-1)I\mathfrak{w}  & \dots & -\mathfrak{w}I \\ \vdots & \ddots & \vdots\\ -\mathfrak{w}I & \dots & \mathcal{L}_{\frac{1}{2}(A^k+(A^k)^T)} -(k-1)I \end{pmatrix} \bar{s} \\
&= \sum_{\alpha =1}^k \left( s_\alpha^T \mathcal{L}_{\frac{1}{2}(A^\alpha+(A^\alpha)^T)}s_\alpha + (k-1)\mathfrak{w}s_\alpha^Ts_\alpha \right ) -\sum_{\alpha=1}^k \sum_{\beta \neq \alpha} \mathfrak{w} s_\alpha^T s_\beta \\
 &=\sum_{\alpha =1}^k s_\alpha^T \mathcal{L}_{\frac{1}{2}(A^\alpha+(A^\alpha)^T)}s_\alpha +k^2n\mathfrak{w}- \mathfrak{w} \sum_{\alpha,\beta =1}^k s_\alpha^Ts_\beta.
 \end{split}
 \end{equation*}
Considering cases when the number of layers becomes large and the layers are relatively sparse demonstrates a consequence of using this approach --- the second and third terms in the expression dominates and the clustering focuses on the cliques of node copies.  But, to make those terms as small as possible we need $s_\alpha$ and $s_\beta$ to be as similar as possible.  In other words, copies of a single node should have the same indicator vector values.  On the other hand, if $\mathfrak{w}$ is relatively small, we can make the first term vanish by making the $s_\alpha$ a multiple of $\vec{1}$.  If that type of choice minimizes the cut, then spectral clustering returns clusters which are simply groups of layers.  The tension between the first and last two terms shows us that copies of a single node will be placed together unless the contribution of the intra-layer connectivity is sufficiently high compared to the inter-layer weights.

Performing the same computation with the multiplex dynamical framework is somewhat more complex.  If we write $\mathcal{L}_\mathfrak{A}=\mathfrak{D}-\mathfrak{A}$, then $\mathfrak{D}$ is a block diagonal matrix where $\mathfrak{D}^{\alpha,\alpha}_{ii}=\frac{1}{2} \left ( \sum_\beta\sum_j B^{\alpha,\beta}_{ij}+\sum_\beta\sum_i B^{\beta,\alpha}_{ji} \right )$ and $B^{\beta,\alpha}=C^{\beta,\alpha}A^\alpha+ (C^{\alpha,\beta} A^\beta)^T$.  So, $\mathfrak{D}^{\alpha,\alpha}=\sum_\beta \frac{1}{2}(deg_{in}^{\alpha,\beta}+deg_{out}^{\beta,\alpha})$ where $deg^{\alpha,\beta}_{in}$ is the diagonal matrices of in-degrees of the matrix $B^{\alpha,\beta}$ and $deg_{out}^{\alpha,\beta}$ is the analgous diagonal matrix of out-degrees.  So, we have
\begin{equation*}
\begin{split}
\bar{s}^T \mathcal{L}_\mathfrak{A} \bar{s} &=\sum_{\alpha,\beta =1}^k s_\alpha^T (\mathfrak{D}^{\alpha,\beta} -B^{\alpha,\beta}) s_\beta \\
&= \sum_{\alpha}^k s_\alpha^T \mathfrak{D}^{\alpha,\alpha} s_\alpha -\sum_{\alpha,\beta=1}^k s_\alpha^T B^{\alpha,\beta} s_\beta \\
&= \sum_\alpha s_\alpha^T \mathcal{L}_{B^{\alpha,\alpha}} s_\alpha + \frac{1}{2}\sum_{\alpha \neq\beta} s_\alpha^T (deg_{in}^{\alpha,\beta}+deg_{out}^{\alpha,\beta})s_\alpha  -\sum_{\alpha \neq \beta} s_\alpha^T B^{\alpha,\beta}s_\beta\\
&= \sum_\alpha s_\alpha^T \mathcal{L}_{B^{\alpha,\alpha}} s_\alpha + \sum_{\alpha \neq \beta} \left ( M^{\alpha,\beta} - s_\alpha^T B^{\alpha,\beta}s_\beta \right ),
\end{split}
\end{equation*}   
where, in the last line, $M^{\alpha,\beta}$ is the total weight of the edges given by $B^{\alpha,\beta}$.  As with the analysis of the supra-Laplacian operator, we isolate the terms with the layer graph Laplacians so we can more clearly see the effect of inter-layer interaction.  The summand in the last term is the total weight of the edges in $B^{\alpha,\beta}$ discounted by by the edge weights between nodes in layer $\alpha$ and layer $\beta$ that are in the same cluster.  Here we see a similarity with the computation using the supra-Laplacian - the tension between the intra- and inter-layer connectivity drives the clustering problem.  However, in this case, there is no parameter akin to the weight $\mathfrak{w}$ in the supra-Laplacian.  Instead, the choices of the $C^{\alpha,\beta}$ serve a similar purpose.  As we will see below, despite their similarities the two sets of modeling choices provide different outcomes for the problem of partitioning the multiplex network.

\section{Clustering multiplex networks with Erd\H{o}s-R\'enyi layers} \label{sec:ER}
In our first set of experiments, we generate multiplex networks where each layer is an Erd\H{o}s-R\'enyi (ER) network with the same fixed wiring probability $p$.   As ER networks generically have no community structure, we expect no structure to arise from an application of spectral clustering.  In fact, we use this case to understand how the presence of the layers themselves differentiate the multiplex networks from the single layer networks through the lens of efficient partitioning.  As the supra-adjacency formulation includes explicit intra-node connections, we expect that if the number of layers outstrips the expected degree of the node copies in each individual layers, the node copies should be clustered together.  Similarly, if the density within the layers is sufficiently high compared to the number of layers, we expect the layers to separate from one another in the clusters.  

In the multiplex dynamical framework,  we expect the community structure is less sensitive to the number of layers.  The neighborhoods of nodes in each layer is, in part, mirrored across other layers through the off-diagonal blocks in $\mathcal{A}$.  Consequently, separating node copies from one another can be more costly than in the supra-adjacency model.  

\begin{figure}
\centering
\includegraphics[width=3in]{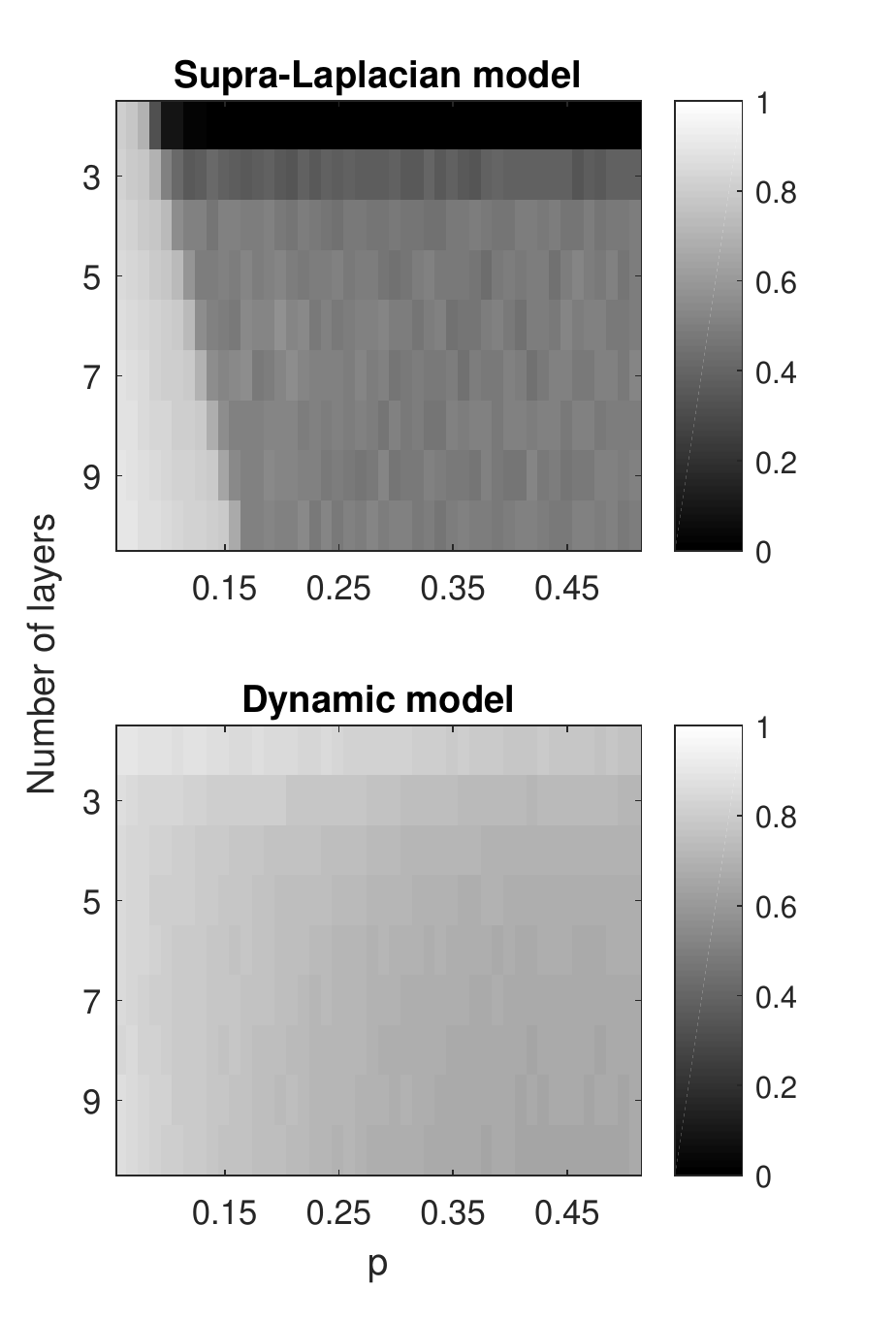}

\caption{Each part of the figure shows the fraction of node copies grouped together over different choices of connection probability $p$ and number of layers $k$.  The top figure shows the results for the spectral clustering using the supra-Laplacian while the bottom shows the spectral clustering using the graph Laplacian associate to our second structural model.}\label{fig:ERlayer}
\end{figure}

To test this, for each $p \in \{0.05,0.06,\dots, 0.5\}, k \in \{2,\dots,10\}$, we create $100$ multiplex networks with $k$ ER layers, each with wiring probability $p$.  For each instance, we perform spectral clustering with two clusters on both the supra-adjacency ($\mathfrak{w}=1$) and multiplex dynamics formulations ($C^{\alpha,\beta}=I$).  We further calculate the mean of the fraction of node copies that are grouped together in the clustering over all instances.  Figure \ref{fig:ERlayer} shows the results, which confirm our conjectures.  The top panel shows that as $p$ increases, the  node copies increasingly do not cluster together when using the supra-Laplacian.  This is most transparent in the case when there are two layers.  For $p$ greater than roughly $0.1$ the layers split, one in each cluster.  Similarly, for larger values of $k$ and $p$ sufficiently large, layers do not split across clusters but land entirely inside one or the other.  Consequently the mean fraction of node copies grouped together is approximately one half.  For smaller enough values of $p$, the partition determined by the eigenvector of the supra-Laplacian does group most of the node copies together, but only because one of the two clusters is very large compared to the other.  

For the multiplex dynamical model, node copies are often grouped together.  The mean fraction of node copies grouped together is close to one for small $p$ and decreases as $p$ increases.  The minimum over all parameters testted is roughly $0.65$.  

\section{Clustering multiplex networks with fixed communities}\label{sec:SBM}
Moving to our next set of investigations, we consider synthetic networks with different kinds of fixed communities encoded on the layers.  We think of these in analogue with the results of the stochastic block model in the single layer case --- we plant communities on different layers to test what different versions of spectral clustering detect.  Our approach has similarities to a recent framework for building multi-layer networks with meso-scale structure \cite{Bazzi_2016}.  In that paper, the authors first fix the desired relationships between the clusters on the different layers and then construct the clusters.  In our first test below, the inter-layer relationship is very rigid --- the clusters must be identical.  In the second test below, this relationship is much looser.  We emphasize that the second experiment does not quite fall within the referenced framework as we construct the model networks differently.

We first test a situation where we expect that any reasonable approach to clustering should succeed.  Instead of using ER networks on each layer, we instead construct a network with two communities generated using a stochastic block model (SBM).  For layer $\alpha$, we place the first $50$ nodes in one community and the second $50$ in another with the intra-cluster connection probability set to $1$ and the inter-cluster connection probability $p\in\{0,0.1,\dots,1\}$.  With these structural elements in place, we construct a family of supra-adjacency matrices indexed by the intra-node weight $\mathfrak{w} \in\{0,0.1,\dots,5\}$ as well as the dynamical multiplex model with $C^{\alpha,\beta}=I$.  

Performing spectral clustering using the graph Laplacians associated to the dynamical multiplex models finds the fixed communities every time over $100$ instances for each choice of $p$.  The supra-adjacency formulation does not fare as well, only finding the communities when the intra-node weightings are sufficiently high compared to the inter-community connection probabilities.     
\begin{figure}
\centering
\includegraphics[width=3.5in]{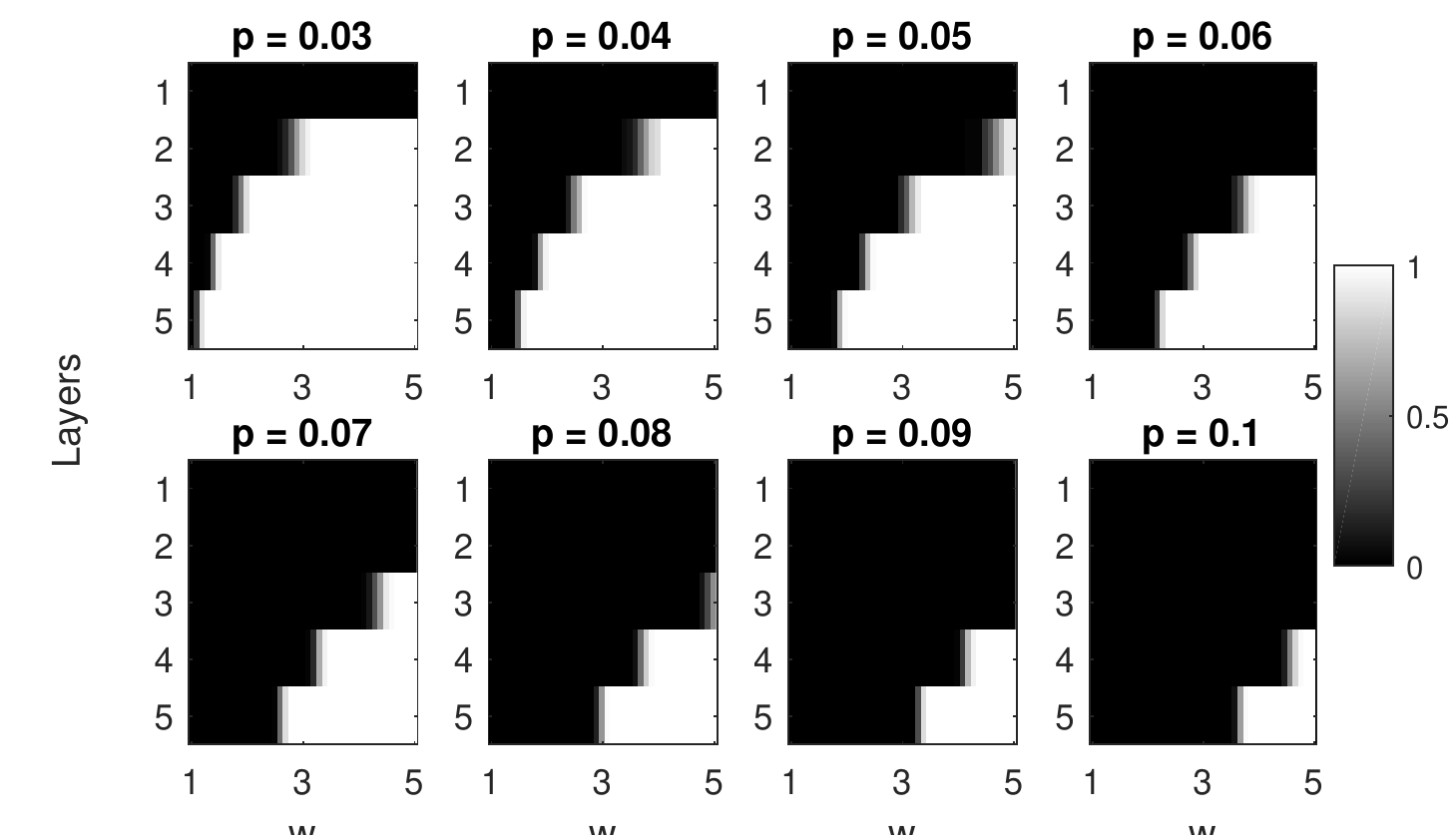}

\caption{Each sub-figure shows the fraction of instances of the experiment where spectral clustering with the supra-Laplacian finds the correct communities over three parameters.  Each sub-figure is associated a different value of $p$, the inter-community connection probability in the SBM, and for that fixed $p$ the experiment is run for different values of $\mathfrak{w}$ (along the x-axes) and the number of layers in the multiplex network.}\label{fig:SBMlayer}
\end{figure}

Figure \ref{fig:SBMlayer} shows the results for several values of $p$ where there are some combinations of $k$ and $\mathfrak{w}$ where spectral clustering with the supra-Laplacian finds the communities correctly.  The results reveal that as the number of layers increase, the supra-Laplacian formulation can detect the communities effectively with smaller weights.  However, as the inter-community connection probabilities increase, accurate identification becomes more and more difficult, requiring higher intra-node weights.  

\begin{figure}
\centering
\includegraphics[width=3.25in]{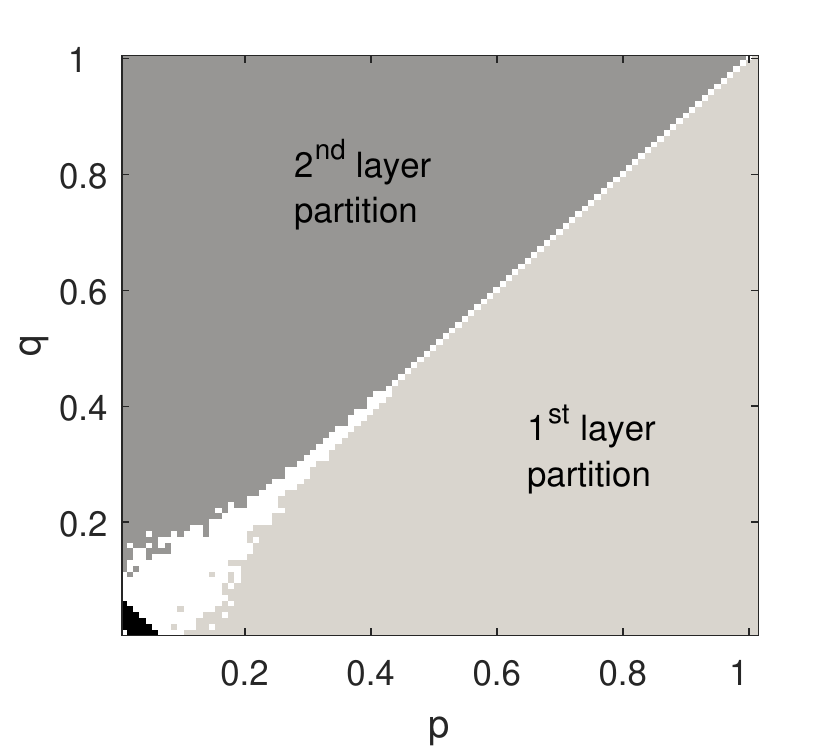}

\caption{This figure shows results of spectral clustering into two clusters on a two-layer network with distinct but overlapping clusters places on each layer using the stochastic block model.  The image shows four regimes:  (black) the clusters are the two layers, (light gray) the clusters correspond to the clusters constructed on the first layer, (dark gray) the clusters correspond to the clusters constructed on the second layer, and (white) the clusters do not correspond to any of the other cases. }\label{fig:SBMoverlap}
\end{figure}

Last, we explore how the dynamical multiplex model deal with an ambiguous case.  For a multiplex network with two layers of $100$ nodes, we encode nodes $\{1,\dots,50\}$ and $\{51,\dots,100\}$ as communities on the first layer and nodes $\{26,\dots,75\}$ and $\{1,\dots,25\}\cup \{76,\dots, 100\}$ as communities on the second layer using the SBM.  We then test spectral clustering on this network using the dynamical multiplex representation with $C^{1,2}=pI, C^{2,1}=qI, C^{1,1}=(1-p)I,$ and $C^{2,2}=(1-q)I$, varying the values of $p$ and $q$.  As summarized in Figure \ref{fig:SBMoverlap}, we find four types of regimes when looking for a partition into two sets over $100$ instances of the problem --- (black) splitting the layers from one another, (light gray) finding the partition given by the first layer, (dark gray) finding the partition given by the second layer, and (white) finding another partition.  In the last case, the partitions are often significantly different for different simulated networks indicating that the results from spectral clustering are delicate in these cases and dependent significantly on other factors in the construction of the SBM. For small values of both $p$ and $q$, clustering separates the two layers from one another.  But, for larger values of $p$ and $q$, if $p>q$ clustering finds the first layer's partition while if $q>p$, it finds the second layer's partition for almost all the instances of the experiment.  When $p$ and $q$ are nearly equal,  clustering results are ambiguous, sometimes finding one or the other or mixtures of the two. 

In a similar experiment, spectral clustering under the supra-adjacency formulation  proves sensitive to very small network perturbations.  We again create a two layer multiplex network with $100$ nodes and the same community structures as in the previous experiment, but with varying inter-cluster probabilities in the SBM.  In building the supra-adjacency matrix, we use weights $\mathfrak{w} \in \{0.1, 0.2,\dots,5\}$.  Applying spectral clustering to this family of networks to find two clusters reveals that for smaller weights, roughly less than $2$, spectral clustering splits the layers away from one another.  For higher weights, the results are similar to the last case in the previous paragraph --- spectral clustering finds other partitions of the network which are neither of the communities generated by the SBM.  Again similarly to the previous experiment, in this case the partitions detected by spectral clustering differ significantly from one another over the $100$ iterations of the experiment for a fixed set of parameters.  

For both of these models, there are regions of the parameter space where the two cluster partition does not align with either of the partitions on the layers created using the SBM.  The ambiguity of these results points to a sensitivity of the methods to the details of the construction which, in turn, may be a consequence of the choice of the number of clusters we use in spectral clustering.  If we think about how the two partition structures on the layers may interact, we see two reasonable possibilities.  First, a method might find two clusters on each layer, for a total of four clusters, by splitting the layers apart and then splitting each layer according to the results of the SBM.  Second, a method may mix the two partitions together and find clusters associated to intersections.  This would lead to four clusters if the node copies are grouped together, eight clusters if they are separated, and various other possibilities if the clusters are mixed together.  To better understand this region of the parameter space, we re-test the synthetic networks with higher numbers of clusters using unormalized spectral clustering for more than two clusters (see Section 4 in \cite{vonLuxberg}).    When using the supra-Laplacian, we find that when the weight parameter is between roughly $2$ and $25$, spectral clustering with four clusters finds the same clustering in every instance.  The partitioning splits the layers apart and finds the two clusters on each layer that arise from the SBM.  At higher weight values, spectral clustering consistently finds $3$ clusters which are mixtures of the two layers clusterings.  Similarly, when $p=q$ in the dynamic multiplex model spectral clustering also finds $3$ clusters formed from intersections of the layer clusterings.

\section{Conclusion}
Our investigation into extending the ideas of spectral clustering to the multiplex network setting points us to an important basic question that separates clustering multiplex networks from clustering in simpler single layer networks --- how do we treat nodes across the layers?  At the two ends of this spectrum --- where nodes are indivisible and where node copies are entirely independent --- multiplex spectral clustering reduces to spectral clustering in an appropriately constructed single layer network.  

For the cases in between, our analysis of two models provides different results related to choices in model construction.  We find that a structural representation based on a multiplex dynamic model \cite{JCN} is a good candidate for a general-purpose model for spectral clustering.   This model performed well in our first two clustering experiments (Sections \ref{sec:ER} and \ref{sec:SBM}) over an array of parameter choices.  This robustness across the parameter space indicates that practitioners may use parameters appropriate to their applications rather than needing to tune parameters to ensure good clustering.  The results of the third experiment demonstrate that in the presence of complex layer interactions, where different choices in the parameter space lead to different partitions arising from spectral clustering.   This is appropriate as the parameters in this model reflect the relative importance of different layers in the context of whatever questions are under investigation.  However, when the parameters linking the two layers are roughly equal results of spectral clustering are ambiguous, demonstrating the need for careful consideration in  this case.   Practitioners need to closely examine these results of spectral clustering as small changes in the network construction can lead to different partitions.   

We see two related results of spectral clustering using the supra-adjacency representation \cite{Gomez}.  First, success in the experiments we evaluate depends on the strength of the inter-layer weight parameter --- different magnitudes lead to different results.  If the weight is low, then spectral clustering has the tendency to split the layers into different clusters.  When the weight is high enough, clustering can recover identical fixed communities across the layers.  But at similar weights when there are different communities on the layers, spectral clustering splits the layers apart.  Only when the weight is much higher does spectral clustering result in partitions which mix together the layer partitions in ways similar to the ambiguous cases we find in the other model.  As shown in \cite{Gomez}, when the weight parameter is sufficiently high, the spectrum of the supra-Laplacian is linked to the spectrum of the average aggregate of the layers.  This leads to our second observation, that using the supra-Laplacian for spectral clustering breaks coarsely into two regimes --- when the inter-layer weight is small we are essentially partitioning into layers and clustering there while when the weight is large enough we are clustering on a single aggregate network.  This reflects the modeling choices inherent in the weight parameter --- low values indicate a weak connection between the layers while large values knit the layers together more and more tightly.  Consequently, we see the supra-adjacency formulation as an appropriate structural model for spectral clustering in cases where the weight parameter is a meaningful reflection of the role of node copies in the network.

\nocite{*}

\section*{Acknowledgments} 

\section*{References}
  \bibliographystyle{elsarticle-num} 
  \bibliography{sc}
\end{document}